\documentclass[twocolumn,showpacs,preprintnumbers,amsmath,amssymb,superscriptaddress]{revtex4}

\usepackage{graphicx}
\usepackage{dcolumn}
\usepackage{bm}

\newcommand{\mnras}{Mon. Not. R. Astron. Soc.}
\newcommand{\aap}{Astron. Astrophys.}
\newcommand{\physrep}{Phys. Rep.}
\newcommand{\araa}{Ann. Rev. Astron. Astrophys.}
\begin{document}


\title{Is the PAMELA anomaly caused by the supernova explosions \\ near
the Earth?}

\author{Yutaka Fujita}
\affiliation{%
Department of Earth and Space Science, Graduate School of Science, Osaka
 University, Toyonaka, Osaka 560-0043, Japan
}

\author{Kazunori Kohri}%
\affiliation{%
Department of Physics, Tohoku University,
Sendai 980-8578, Japan
}%
\affiliation{%
Physics Department, Lancaster University, Lancaster LA1 4YB, UK
}%

\author{Ryo Yamazaki} \affiliation{ Department of Physical Science,
Hiroshima University, Higashi-Hiroshima, Hiroshima 739-8526, Japan }

\author{Kunihito Ioka}
\affiliation{
KEK Theory Center and the Graduate University for Advanced Studies
(Sokendai), 1-1 Oho, Tsukuba 305-0801, Japan
}%

\date{\today}

\begin{abstract}
 We show that the anomaly of the positron fraction observed by the
 PAMELA experiment can be attributed to recent supernova explosion(s) in
 a dense gas cloud (DC) near the Earth. Protons are accelerated around
 the supernova remnant (SNR). Electrons and positrons are created
 through hadronic interactions inside the DC. Their spectrum is harder
 than that of the background because the SNR spends much time in a
 radiative phase. Our scenario predicts that the anti-proton flux
 dominates that of the background for $\gtrsim 100$~GeV. We compare the
 results with observations (Fermi, HESS, PPB-BETS, and ATIC).

\end{abstract}

\pacs{Valid PACS appear here}
\maketitle

\section{Introduction}

Observations with the PAMELA \cite{adr08} and ATIC/PPB-BETS \cite{cha08}
experiments have implied or shown anomaly in the positron fraction and a
sharp feature in the CR electron spectrum, although Fermi satellite
recently reported that the feature in the electron spectrum is much
smaller or even absent depending on the uncertainty of the background
cosmic-ray electron and positron spectrum \cite{Abdo:2009zk}. The PAMELA
(and ATIC/PPB-BETS) results indicate that there should be an unknown
source of electrons and positrons. One attractive idea of the source is
the annihilation or decay of weakly interacting dark matter particles
\cite{his05,mal09}. The other is astrophysical scenarios such as pulsars
\cite{mal09,chi96,Kawanaka:2009dk}.

Among the astrophysical sources, supernova remnants (SNRs) are one of
the promising candidates for the electron and positron source. Their
synchrotron emission reveals the existence of high energy electrons in
SNRs \cite{koy95}, and their $\gamma$-ray emission would indicate that
of high energy protons \cite{aha04}, although there has been no direct
evidence of proton acceleration in SNRs yet.

In this paper, we show that the observed electron and positron excesses
can naturally be explained by recent supernova explosions near the
Earth. In fact the low-density local bubble (LB), in which the solar
system locates, and the adjacent bubble (Loop~I) with sizes of
$\sim$100~pc are often thought to be created through $\sim$20--40
supernova explosions that started $\sim 10^7$~yrs ago
\cite{ber02}. Since those explosions occurred close to us ($\sim
100$--200~pc), they could have affected the environment around the solar
system more significantly than the supernova explosions in the spiral
arms $\sim$kpc away \cite{sha09}.

\section{Supernova(e) in a Dense Cloud}

We consider supernova explosions that happened $\sim 10^{5-6}$~yrs ago
in a dense gas cloud (DC) $\sim$100--200~pc away from the Earth. The DC
does not necessarily correspond to the host DCs of the supernovae that
created the LB or Loop~I. Massive stars that explode as supernovae tend
to be born in giant DCs \cite{lar82}. They may explode near or inside
the host DCs, because their durations of life are short. In this paper,
we assume that although a DC had not been destroyed by ultra-violet
radiation from massive, short-lived OB stars, it is ionized and the
temperature is $\sim 10^4$~K when those stars explode \cite{whi79}. In
such environments, the shock front of an SNR accelerates protons, which
create electrons and positrons through hadronic interactions in the
surrounding DC.

The evolution of a SNR and the particle acceleration around it is
modeled following a previous study \cite{yam06}. The shock velocity
$v_s$ is written as a function of the SNR age ($t_{\rm age}$) as
\begin{equation}
v_s(t_{\rm age})=
\left\{
\begin{array}
{l@{} l@{}}
v_i & (t_{\rm age}<t_1) \\
v_i \left(\frac{t_{\rm age}}{t_1}\right)^{-3/5} 
& (t_1<t_{\rm age}<t_2) \\
v_i \left(\frac{t_2}{t_1}\right)^{-3/5}
\left(\frac{t_{\rm age}}{t_2}\right)^{-2/3}~ & 
(t_2<t_{\rm age})
\end{array} \right.~~ ,
\label{eq:Vs}
\end{equation}
where $v_i=v_{i,9}10^9$~cm~s$^{-1}$ is the initial velocity, and $E_{\rm
SN}=E_{51}10^{51}$~erg is the initial energy of the ejecta. When we
consider multiple supernova explosions, we simply assume that they
explode almost simultaneously (within a time-scale of the destruction of
the DC) and give a larger $E_{\rm SN}$ than that of a single
supernova. At $t_{\rm age}<t_1 =45\:
(E_{51}/n_2)^{1/3}v_{i,9}^{-5/3}$~yr, the SNR is in the free expansion
phase, at $t_{\rm age}>t_2=3.5\times 10^3 E_{51}^{4/17}n_{\rm
2}^{-9/17}$~yr, it is in the radiative phase, and at $t_1<t_{\rm
age}<t_2$, it is in the Sedov phase. Here $n_0= n_{\rm 2}\:10^2\:\rm
cm^{-3}$ is the proton density of the DC.

We assume that the energy spectrum of the accelerated protons is 
\begin{equation}
\label{eq:NE}
 N_p(E)\propto E^{-s}\exp(-E/E_{\rm max, p})\; .
\end{equation}
The index is represented by $s=(r+2)/(r-1)$, where $r$ is the
compression ratio of the shock \cite{bla87}, which is given by the
Rankine-Hugoniot relation for $t_1<t_{\rm age}<t_2$, and $r\sim
\sqrt{2}v_s/v_{\rm Au}$ for $t_{\rm age}>t_2$, where $v_{\rm Au}$ is the
upstream Alfv\'en velocity \cite{yam06}. The maximum energy is
determined by the age of SNR:
\begin{equation}
 E_{\rm max,p}=1.6\times10^2~h^{-1} v_{s,8}^2
\left(\frac{B_{\rm d}}{10~\rm\mu G}\right)
\left(\frac{t_{\rm age}}{10^5{\rm yr}}\right)~{\rm TeV}~,
\label{eq:Emax_p}
\end{equation}
where $v_{s,8}=v_s/10^8$cm~s$^{-1}$, $h~(\sim 1)$ is the factor
determined by the shock angle and the gyro-factor, and $B_{\rm d}$ is
the downstream magnetic field, which is expressed by $B_{\rm d}=rB_{\rm
DC}$, where $B_{\rm DC}$ is the magnetic field in the DC \cite{yam06}.
Since we consider an ionized DC, we ignore the ion-neutral wave damping
at the upstream of a shock (for particle acceleration in a radiative
phase, see \cite{dru96}). We equate the minimum energy of the protons
with the proton rest-mass energy.

We assume that supernovae explode at the center of a DC for the sake of
simplicity. We take $v_{i,9}=1$, $h=1$, and $B_{\rm DC}=10\rm\:
\mu G$ from now on.

We consider the case where a few supernovae (or one hypernova) explode
almost simultaneously. Thus we assume that $E_{51}=3$, from which we
obtain $t_1=65$~yr and $t_2=4.5\times 10^3$~yr, if the proton density
inside the DC is $n_0=100\:\rm cm^{-3}$. At $t_{\rm age}=8.0\times
10^4$~yr, the Mach number and radius of the SNR are $M_a=7$ and
$R=26$~pc, respectively. At this point, the SNR is in the radiative
phase. We assume that particle acceleration around a SNR ceases for
$M_a\lesssim 7$ (see e.g. \cite{ryu03}), although the exact value is not
clear. We refer to the age at which $M_a=7$ as $t_{\rm age}=t_{\rm
acc}$.

At $t_{\rm age}=t_{\rm acc}$, we obtain $s=1.0$ and $E_{\rm max, p}\sim
120$~TeV. Although particle acceleration stops at this time, the SNR
continues to expand. The radius is $R_s=50$~pc at $t_{\rm age}=5\times
10^5 $~yr. Since it is comparable to the size of a giant DC \cite{mck07}
and $E_{\rm SN}$ is larger than the binding energy of a DC, the cloud
would be destroyed around this time. Until destroyed, the DC is
illuminated by the accelerated protons from the inside with the spectrum
of Eq. (\ref{eq:NE}) given at $t_{\rm age}\sim t_{\rm acc}$. The
duration of the exposure ($t_{pp}$) could be approximated by the time
elapsing from the explosion of the supernovae to the destruction of the
DC because of the short time-scale of $t_{\rm acc}$. The protons create
$e^{-}$, $e^{+}$, $p$ and $\bar{p}$ through $pp$ collisions in the DC
for the period of $\sim t_{pp}$. Based on this scenario, we calculate
their spectrum observed at the Earth, and search parameters that are
consistent with observations. Since the model of the SNR evolution may
have some uncertainties, we do not stick to the exact values of
parameters in the above discussion.

After the destruction of the DC, the created $e^{-}$, $e^{+}$, $p$ and
$\bar{p}$ propagate through diffusion processes and reach to the
Earth. Since we assume that the DC has already been destroyed until the
current epoch, there is a difference in the last arrival time between
massless neutral particles such as photons and neutrinos, and charged
particles. We would not detect any photon and neutrino signals from the
DC $\sim 10^{5-6}$ yrs after the destruction.

We have calculated spectra of those daughter particles of the $pp$
collisions by performing the PYTHIA Monte-Carlo event
generator~\cite{Sjostrand:2006za}. Then we solve the diffusion
equation of the charged particle ``$i$'' ($i$ runs $e^{-}$, $e^{+}$,
$p$ and $\bar{p}$),
\begin{equation}
    \label{eq:diff_eq}
    \frac{\partial f_{i}}{\partial t} 
= K(\varepsilon_{i})\nabla^{2}f_{i} +
    \frac{\partial}{ \partial \varepsilon_{i}} \left[
    B(\varepsilon_{i}) 
f_{i}\right] + Q(\varepsilon_{i})
\end{equation}
where $f_{i}(t,{\boldmath{x},\varepsilon_{i}})$ is the distribution
function of $i$ and $\varepsilon_{i} = E_{i}/{\rm GeV}$ with $E_{i}$
being the energy of $i$.  The flux $\Phi_{i}
(t,{\boldmath{x},\varepsilon_{i}})$ is given by $\Phi_{i} = (c/4\pi)
f_{i}$. For electrons and positrons, according to a model 08-005
in~\cite{Moskalenko:1997gh}, we take
\begin{equation}
 K(\varepsilon_{e}) = K_{0}(1 +
\varepsilon_{e}/3{\rm GeV})^{\delta}
\end{equation}
with $K_{0} = 2 \times 10^{28}
{\rm cm}^{2}{\rm s}^{-1}$ and $\delta = 0.6$. The cooling rate through
the synchrotron emission and the inverse Compton scattering is given by
\begin{equation}
 B(\varepsilon_{e}) = 10^{-16} {\rm s}^{-1} \varepsilon_{e}^{2 }[ 0.2
(B_{\rm diff}/3 \mu {\rm G})^{2} + 0.9 ]/1.1
\end{equation}
\cite{Baltz:1998xv}, where $B_{\rm diff}$ is the magnetic field outside
the DC. These parameters approximately correspond to so-called ``med
model'' of the particle propagation~\cite{Bottino:2005xy}.

If we assume that the timescale of the production is shorter than the
diffusion timescale $\sim d^{2}/(Kc)$, and the daughter-particle source
is spatially localized well, we can also use the analytical solution
in~\cite{Atoian:1995ux}.  When the shape of the source spectrum is a
power-law with an index $\alpha$, $Q =
Q_{0}\varepsilon^{-\alpha}\delta(\boldmath{x})\delta(t)$, the solution
is represented by
\begin{equation}
    \label{eq:f_diff}
f_{e}=\frac{Q_0 \varepsilon_e^{-\alpha}}{\pi^{3/2} d_{\rm diff}^3}
\left(1-\frac{\varepsilon_e}{\varepsilon_{\rm cut}}\right)^{\alpha-2}
e^{-(\bar{d}/d_{\rm diff})^2},
\end{equation}
with $\varepsilon_{\rm cut} = (B t_{\rm diff})^{-1}$ and the diffusion
length 
\begin{equation}
 d_{\rm diff} = 2 \sqrt{K t_{\rm diff} \frac{1 - (1 -
\varepsilon_e/\varepsilon_{\rm cut})^{1 - \delta}}{
(1-\delta)\varepsilon_e/\varepsilon_{\rm cut}}}\:.
\end{equation}
Here we took an effective distance to the source $\bar{d}$, which is
calculated by spatially averaging the distance to the volume element of
the source, and $\alpha \simeq s$ in the current calculation. The time
after the input of the daughter particles is defined by $t_{\rm
diff}$. Approximately $Q_0 \varepsilon_{i}^{-\alpha}$ is represented by
$\sim V_{s}t_{pp} d^{2}n_{i}/(dtdE_{i})$ with $V_{s}$ the volume of the
source and 
\begin{equation}
 \frac{d^{2}n_{i}}{dtdE_{i}} = \int d E_{p} n_{0} N_{p} \sum_{j}
g_{j} c \frac{d\sigma_{j}}{dE_{i}} \:.
\end{equation}
Here the differential cross section of the ``$j$''-mode for the
production of $i$ particle is $d\sigma_{j}(E_{p},E_{i})/dE_{i}$ with its
multiplicity $g_{j} = g_{j}(E_{p},E_{i})$. We also consider the neutron
decay for the electron/positron production, which is not included in the
original version of PYTHIA. The initial proton spectrum $ N_{p}(E_{p})$
is normalized to be 
\begin{equation}
 V_{s}\int dE_{p} N_{p}(E_{p}) = E_{\rm tot, p} \:,
\end{equation}
which is given later.

For the local propagation of $p$ and $\bar{p}$, the cooling term is
negligible.  In addition, we may omit the annihilation of antiprotons
through scattering off the background protons because the mean free path
for the annihilation reaction is large. We can also omit the convection
by the convective wind. Thus the analytical solution for the proton and
antiproton can be also given by Eq.~(\ref{eq:f_diff}) with taking a
limit of $\varepsilon_p/\varepsilon_{\rm cut} \to 0$.

\section{Results}

\begin{figure}
\includegraphics[width=90mm]{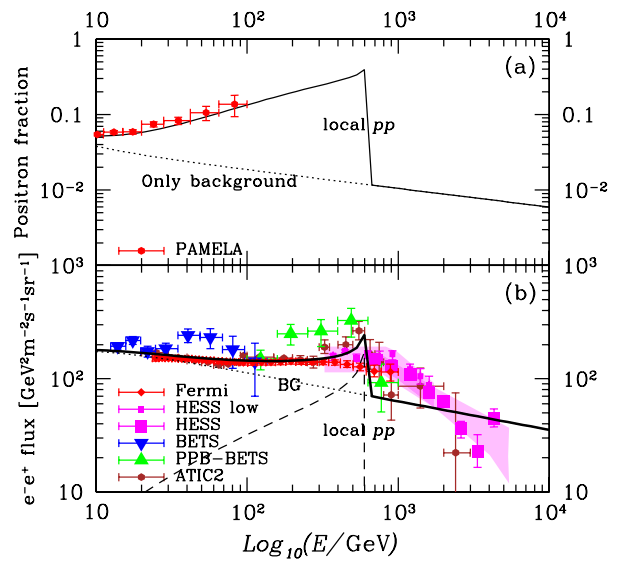} 
\vspace{-1.8 cm}
 \caption{(a) Positron fraction for Case~A (solid line), which includes
 the electrons and positrons coming from the DC and background electrons
 (dotted line). Filled circles correspond to the PAMELA data
 \cite{adr08}. (b) Electron and positron flux for Case~A (solid
 line). The flux of the electrons and positrons created in the DC is
 shown by the dashed line, and that of the background ones is shown by
 the dotted line. Observational results (Fermi, HESS, BETS, PPB-BETS,
 and ATIC2 \cite{cha08,Abdo:2009zk,Aharonian:2009ah}) are also
 presented. The shadow shows the uncertainty of the HESS data. }
 \label{fig:fr1}
\end{figure}

Fig.~\ref{fig:fr1} shows the positron fraction and the $e^-$-$e^+$ flux
observed at the Earth. We assume that a spherical DC with a radius of
$R_{\rm DC}=40$~pc and proton density of $n_0 = 100\rm\: cm^{-3}$ is
uniformly illuminated by protons with the spectrum given by
Eq. (\ref{eq:NE}) with $s=1.4$ and $E_{\rm max}=100$~TeV for the
duration of $t_{pp}=5\times 10^5$~yr. The total energy of the primary
cosmic-ray protons is taken to be $E_{\rm tot,p}=3\times
10^{50}$~erg. The distance to the front side of the DC is $d=200$~pc. We
assume that the diffusion time of $e^-$ and $e^+$ is $t_{\rm
diff}=5\times 10^5$~yr, and $B_{\rm diff}=3\rm\: \mu G$. We refer to
this parameter set as Case~A. Observational results are also shown. This
result is consistent with observations for the positron fraction, and
the $e^-$-$e^+$ flux obtained with PPB-BETS and ATIC2.

The feature of the positron spectrum can intuitively be understood as
follows. The spectrum of the secondary $e^-$ and $e^+$ from the DC is
harder than that of the background one, because the primary protons
are mainly produced in the radiative phase of the SNR with the
compression ratio $r$ much larger than 4. The cutoff scale is given by
$E_{\rm cut} = (B t_{\rm diff})^{{-1}}$ GeV $\sim 600\: {\rm GeV}/
(t_{\rm diff}/5\times10^{5}{\rm yrs})$. The sharp rise around the cutoff
energy is made by $e^-$ and $e^+$ accumulated through cooling. That is
because we are considering the case of $s< 2$, which means that many of
the cooled electrons sizably contribute to modifying the spectral shape
around the cutoff energy. However, the actual spectrum would be broader
with a width of $\Delta E_{\rm cut}/E_{\rm cut} \sim t_{pp}/t_{\rm diff}
\sim 1$, because the injection continues for a finite duration $t_{pp}$
\cite{Kawanaka:2009dk}.

\begin{figure}
\includegraphics[width=90mm]{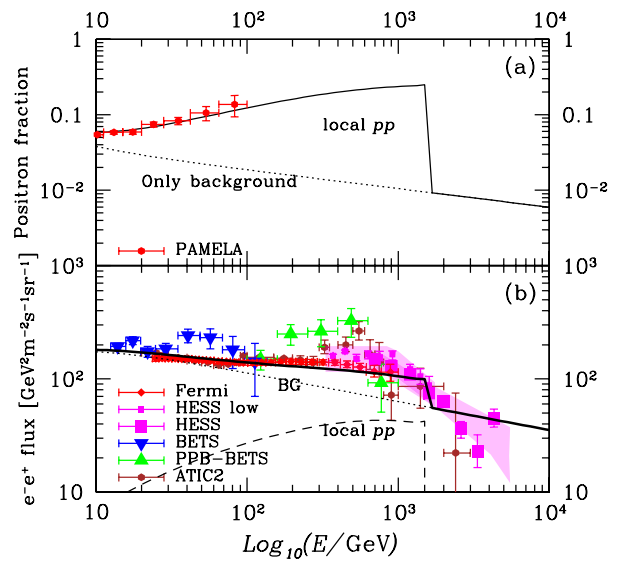} 
\vspace{-1.8 cm}
\caption{Same as Fig.~\ref{fig:fr1}
 but for Case~B.}  \label{fig:fr2}
\end{figure}

\begin{figure}
\vspace{-.2 cm}
    \includegraphics[width=84mm]{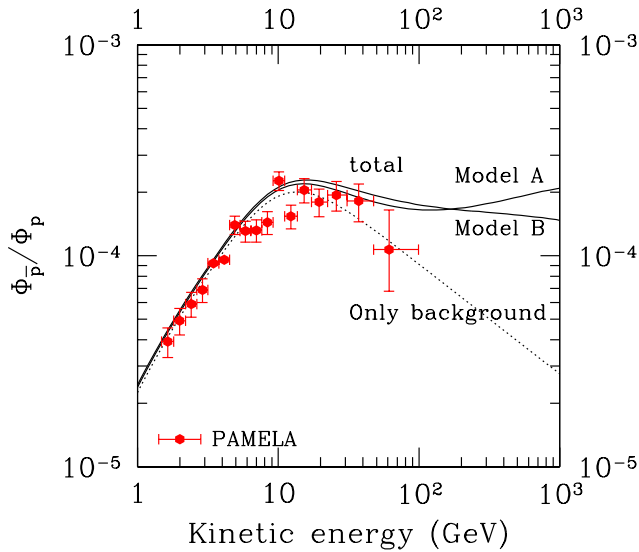}
\vspace{-.2 cm}
 \caption{Antiproton fraction for
    Cases~A and B (solid lines). The dotted line shows the fraction
    calculated only from the background flux.  The fitting formula of
    the background proton and antiproton are taken
    by~\cite{Nezri:2009jd}. We adopted data points taken with
    PAMELA~\cite{Adriani:2008zq}.}  \label{fig:anti_p}
\end{figure}

Fig.~\ref{fig:fr2} is another example of the positron fraction and the
$e^-$-$e^+$ flux. In this case (Case~B), we adopt $R_{\rm DC}=40$~pc,
$n_0=50\rm\: cm^{-3}$, $s=1.75$ , $E_{\rm max}=100$~TeV, $t_{pp}=2\times
10^5$~yr, $E_{\rm tot,p}=3\times 10^{50}$~erg, $d=200$~pc, $t_{\rm
diff}=2\times 10^5$~yr, and $B_{\rm diff}=3\rm\: \mu G$. The results are
consistent with observations for the positron fraction and the
$e^-$-$e^+$ flux obtained with Fermi. Since we have not performed a
complete parameter search, Cases~A and~B should be regarded as examples
among possibilities.

Fig.~\ref{fig:anti_p} shows the antiproton fraction at the Earth. In our
scenario, the flux of anti-protons dominates that of the background for
$\gtrsim 100$~GeV, which can be used to discriminate our hadronic
scenario from other leptonic ones. Antiproton fraction rises at higher
energies than that of the positron fraction (Figs.~\ref{fig:fr1}
and~\ref{fig:fr2}), because the spectral index of the background
antiproton is harder than that of the background positron. So far the
PAMELA antiproton data could not have distinguished one model from
another yet (Fig.~\ref{fig:anti_p}).

We note that old SNRs in DCs could be observed as unidentified TeV
sources \cite{yam06,iok09,fuj09}. Broad-band spectra of the old SNRs
could constrain the parameters of our scenario. It is likely that the DC
has made a star cluster, which might be identified near the Earth.

Recent observations have shown possible decline of B/C ratio toward
higher energies ($\gtrsim 100$~GeV), which seems to be inconsistent with
the present scenario because SNRs also accelerate metals
\cite{CREAM,Mertsch:2009ph}. In our scenario, however, the source of
electrons and positrons is the SNR in a specific dense cloud, and it is
very localized. Thus it is subjected to the variation of metal
abundances in the Galaxy. In fact the abundance of Carbon in Carina
Nebula, which is a star forming region in the Galaxy, is $\lesssim 0.25$
solar value \cite{ham07}. If the dense cloud we considered has such a
low metal abundance, the cosmic-rays from the cloud does not increase
the observed B/C ratio. Moreover our scenario requires that the dense
cloud has not been destroyed by stellar winds before the explosion of
the stars. It is known that stellar winds are weaker for stars with
lower metal abundances, and that the metal abundances of stars are
correlated with those of the host cloud. Thus the cloud in which
electrons and positrons are effectively generated may tend to have low
metal abundances. In summary our scenario cannot be rejected only by the
present observational data of the B/C ratio.

\section*{ACKNOWLEDGMENTS}
We are grateful to referees for valuable comments and P. Blasi,
Y. Ohira, K. Nakayama, and N. Sahu for useful discussions. This work was
supported by KAKENHI, No. 20540269 (Y.~F.), 21740184 (R.~Y.), 19047004
(R.~Y. and K.~I.), and 21684014 (K.~I.). The work of K.~K. was supported
by the European Union through the Marie Curie Research and Training
Network ``UniverseNet'' (MRTN-CT-2006-035863) and by STFC Grant
ST/G000549/1.

\appendix


\end{document}